\documentclass[aps,twocolumn,amsmath,amssymb,prb,showpacs,superscriptaddress]{revtex4-1}
\usepackage{mathtools}
\usepackage{natbib}
\usepackage{array}
\usepackage{xcolor}
\usepackage{subfigure} 
\usepackage{amsmath} 
\usepackage{amsfonts}
\usepackage{amssymb}
\usepackage{graphicx}
\usepackage[bookmarksnumbered,pdfpagelabels=true,plainpages=false,colorlinks=true,linkcolor=blue,citecolor=red,urlcolor=blue]{hyperref}
\usepackage{braket}
\usepackage{cleveref}
\usepackage{comment}
\usepackage{dsfont}

\def\HBdG{{H_{\rm{BdG}}}}
\def\EGS{{E_{\rm{GS}}}}
\def\CalU{{\mathcal{U}}}
\def\CalA{{\mathcal{A}}}
\def\CalB{{\mathcal{B}}}

\begin{document}
  
\title{Steering Majorana braiding via skyrmion-vortex pairs: a scalable platform}

\author{Jonas Nothhelfer}
\affiliation{Faculty of Physics, University of Duisburg-Essen, 47057 Duisburg, Germany}

\author{Sebasti\'an A.\ D\'iaz}
\affiliation{Faculty of Physics, University of Duisburg-Essen, 47057 Duisburg, Germany}

\author{Stephan Kessler}
\affiliation{Institute of Physics, Johannes Gutenberg University of Mainz, 55128 Mainz, Germany}

\author{Tobias Meng}
\affiliation{Institute of Theoretical Physics, Technische Universit\"at Dresden, 01062 Dresden, Germany}
\affiliation{Institute for Theoretical Physics and W\"urzburg-Dresden Cluster of Excellence ct.qmat, Technische Universit\"at Dresden, 01069 Dresden, Germany}

\author{Matteo Rizzi}
\affiliation{Forschungszentrum J\"ulich, Institute of Quantum Control, Peter Gr\"unberg Institut (PGI-8), 52425 J\"ulich, Germany}
\affiliation{Institute for Theoretical Physics, University of Cologne, D-50937 K\"oln, Germany}

\author{Kjetil M. D. Hals}
\affiliation{Department of Engineering Sciences, University of Agder, 4879 Grimstad, Norway}

\author{Karin Everschor-Sitte}
\affiliation{Faculty of Physics, University of Duisburg-Essen, 47057 Duisburg, Germany}
\affiliation{Center for Nanointegration Duisburg-Essen (CENIDE), University of Duisburg-Essen, 47057 Duisburg, Germany}

\date{\today}

\begin{abstract}
Majorana zero modes are quasiparticles that hold promise as building blocks for topological quantum computing. However, the litmus test for their detection, the observation of exotic non-abelian statistics revealed by braiding, has so far eluded experimental efforts. Here we take advantage of the fact that skyrmion-vortex pairs in superconductor-ferromagnet heterostructures harboring Majorana zero modes can be easily manipulated in two spatial dimensions. We adiabatically braid the hybrid topological structures and explicitly confirm the non-abelian statistics of the Majorana zero modes numerically using a self-consistent calculation of the superconducting order parameter. Our proposal of controlling skyrmion-vortex pairs provides the necessary leeway toward a scalable topological quantum computing platform.
\end{abstract}


\maketitle


\section{Introduction}

A major drawback of current quantum computers is the loss of quantum coherence leading to the occurrence and proliferation of errors~\cite{DiVincenzo1995,Landauer1995,Pellizzari1995,Unruh1995}.
Topological quantum computing promises to solve this issue by encoding logical information in anyons with non-abelian statistics~\cite{pachos_2012,Nayak2008,Kitaev2003}.
One type of these topologically stable quasiparticles are Majorana zero modes~\cite{Alicea2012,Leijnse2012,Flensberg2021}, which represent a big step forward in the field of topological quantum computation even though they are not enough for universal quantum computation. Majorana zero modes have been predicted to occur at the ends of a one-dimensional spinless p-wave superconductor~\cite{Kitaev2000}, semiconductor nanowires in proximity to an s-wave superconductor~\cite{Lutchyn2010,Oreg2010} as well as at the core of superconducting vortices in two-dimensional systems ~\cite{Read2000,Ivanov2001,Fu2008,Lutchyn2018}. 
While a few controversial experiments have reported features alluding to the finding of these quasiparticles~\cite{Mourik2012c,Kim2018g}, the ``smoking gun'' experiment, namely, the braiding of two Majorana zero modes revealing their non-abelian statistics, is still missing.

\begin{figure}[tb]
	\includegraphics[width=0.92 \columnwidth]{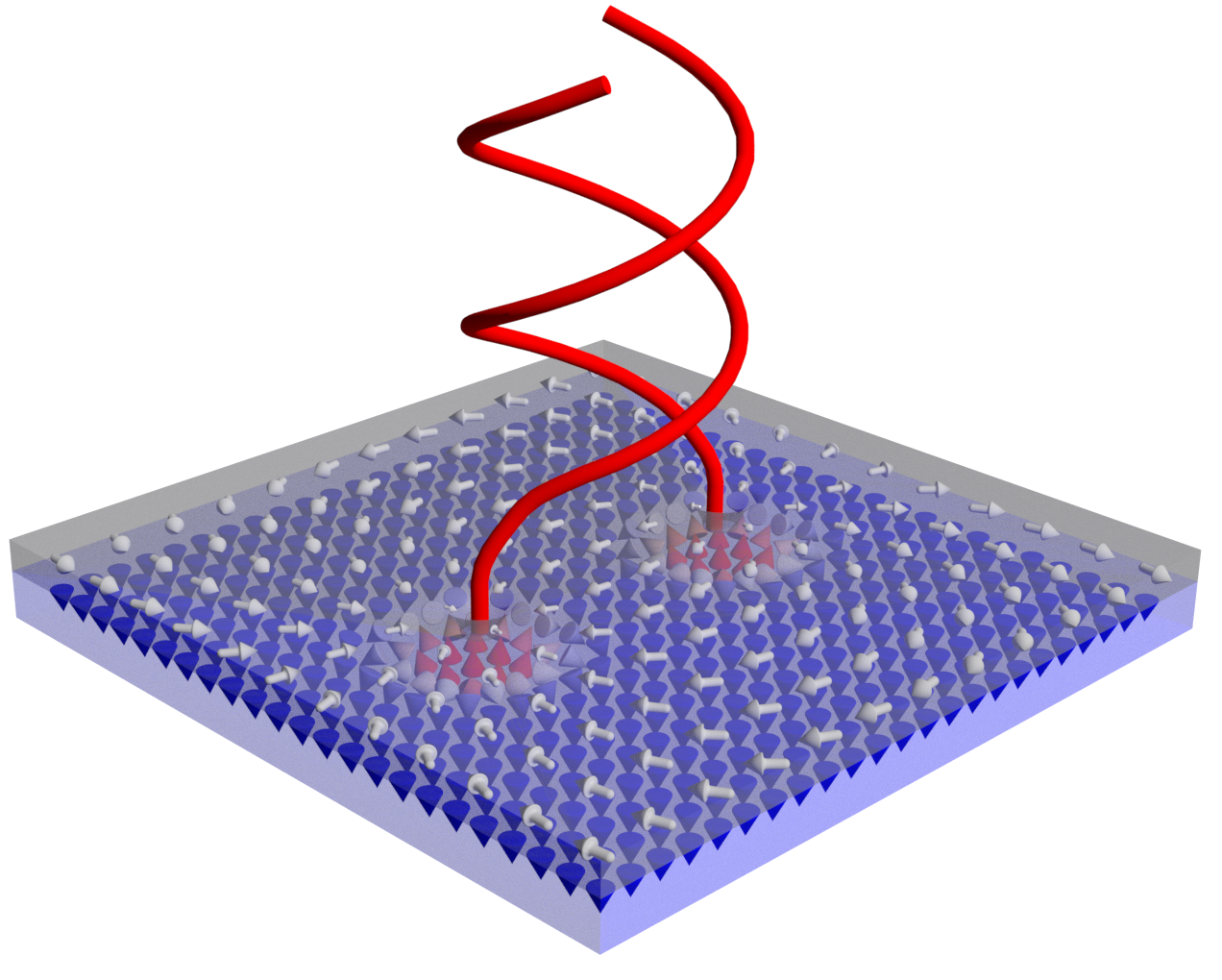}
	\caption{Braiding of two skyrmion-vortex pairs in a superconductor-ferromagnet heterostructure. On the top layer, the length and direction of the white arrows respectively represent the magnitude and phase of the superconducting order parameter. The cones in the bottom layer depict the magnetic structure with upward (downward) pointing magnetic moments plotted in red (blue). The red tubes indicate the world-lines that emerge when braiding Majorana zero modes.}
	\label{fig:SVP}
\end{figure}

There are several theoretical proposals to perform this braiding operation. A famous one employs so-called T-junctions and a suitable manipulation of the chemical potential in quantum wires to realize real-space braiding~\cite{Alicea2011a}.
While T-junctions effectively only allow for the motion of Majorana zero modes in one dimension, Majorana zero modes localized at superconducting vortices might allow for their braiding in two dimensions. However, controlling superconducting vortices in real space without destroying the Majorana zero modes is a challenging enterprise.
Therefore, other concepts have been suggested to demonstrate the non-abelian statistics of Majorana zero modes by performing braiding operations in two-dimensional parameter spaces~\cite{Bonderson2008,Alicea2011a,vanHeck2012,Vijay2016,Karzig2017}. 

Previously, a two-dimensional ferromagnet proximity-coupled to an s-wave superconductor was predicted to stabilize composite topological excitations consisting of a ferromagnetic skyrmion and a superconducting vortex that can be moved by means of spin-torques in the ferromagnet~\cite{Hals2016}. 
These skyrmion-vortex pairs (SVPs) were later shown to host Majorana zero modes~\cite{Rex2019b,Nothhelfer2019}. 
However, whether SVPs and the attached Majorana zero modes remain stable throughout their adiabatic motion has not been so far investigated.

In this work we numerically demonstrate the feasibility of braiding Majorana zero modes attached to SVPs and confirm their non-abelian statistics. A sketch of the braiding of two SVPs in a superconductor-ferromagnet heterostructure is shown in Fig.~\ref{fig:SVP}.
We furthermore show that the braiding operations can still be carried out even when the total number of SVPs in the system is increased. 
Thus, our results are an important step towards a scalable platform for topological quantum computing~\cite{Nothhelferpatent2019}.


\section{Theory of a superconductor proximity-coupled to a ferromagnet }
\label{sec:SFH}

We consider a superconductor-ferromagnet heterostructure, where the two-dimensional s-wave superconductor is modeled by the tight-binding Hamiltonian~\cite{Bjornson2013,Hals2016}
\begin{widetext}
\begin{equation}
H=-t \sum_{\langle ij\rangle} \boldsymbol{c}_i^{\dagger}\boldsymbol{c}_j -\mu \sum_{i}\boldsymbol{c}_i^{\dagger}\boldsymbol{c}_i-\sum_i\boldsymbol{c}_i^{\dagger} (\boldsymbol{h}_i\cdot\boldsymbol{\sigma})\boldsymbol{c}_i
+\mathrm{i} \, \alpha_{\text{R}} \sum_{\langle ij\rangle} \boldsymbol{c}_i^{\dagger}\hat{\boldsymbol{z}}\cdot (\boldsymbol{\hat{d}}_{ij}\times\boldsymbol{\sigma} )\boldsymbol{c}_j +\sum_i (\Delta_ic_{i\uparrow}^{\dagger}c_{i\downarrow}^{\dagger}+h.c.).
\label{eq:TightBindingHamiltonian}
\end{equation}
\end{widetext}
The proximity-coupled ferromagnet is introduced by a magnetic field $\boldsymbol{h}_{i}$ acting on site $i=(x, y)$, which couples to the spins of the electrons in the superconductor. 
Here $\boldsymbol{c}_i^{\dagger}=(c_{i\uparrow}^{\dagger},c_{i\downarrow}^{\dagger})$, where $c_{i\alpha}^{\dagger}$ creates an electron with
spin $\alpha$ at lattice site $i$. The symbol $\langle ij \rangle$ indicates a summation over nearest-neighbor lattice sites, $\boldsymbol{\hat{d}}_{ij}$
is a unit vector that points from site $j$ to site $i$, and $\boldsymbol{\sigma}$ is the vector of Pauli matrices.
The spin-orbit coupling strength $\alpha_\text{R}$, the chemical potential $\mu$, the spatially-dependent superconducting order parameter $\Delta _i$ as well as the magnetic field are measured in units of the the hopping parameter $t =  \frac{\hbar^2}{2ma^2}$, with $a$ being the lattice constant and $m$ the effective mass of the electrons.

\begin{figure*}[tb]
	\includegraphics[width=\textwidth]{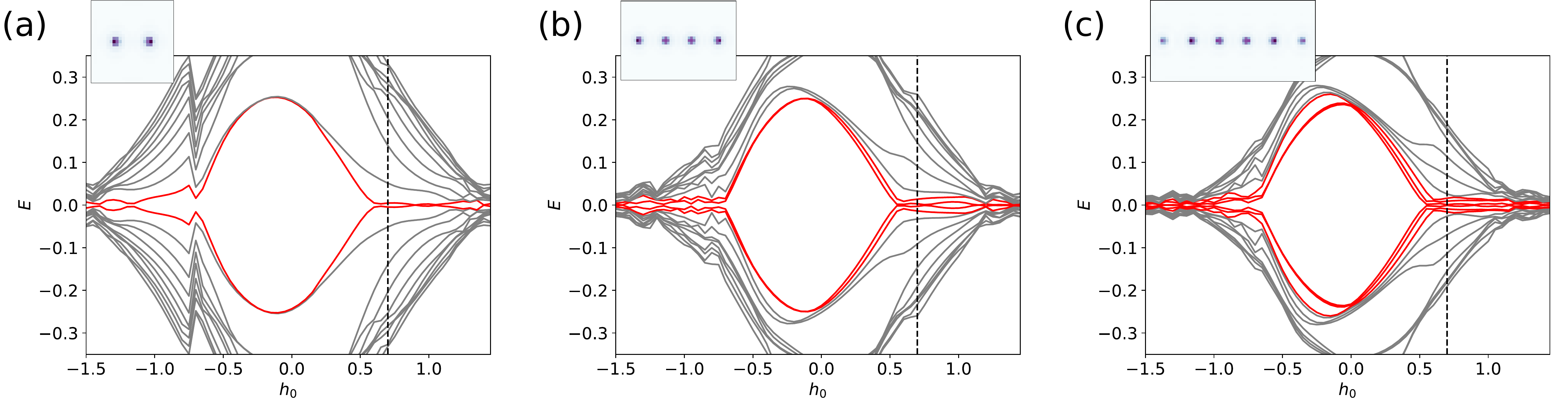}
	\caption{Low energy spectrum of superconductor-ferromagnet heterostructures with (a) two, (b) four, (c) six SVPs as a function of the magnetic field strength $h_0$. For better visibility we marked the modes which experience a gap closing in red. The black dashed vertical line corresponds to $h_0=0.7$, which we study further as in this regime we find localized Majorana zero modes. Notably the number of zero modes corresponds to the number of SVPs.  
	The corresponding insets show the localization of the Majorana zero modes for $h_0=0.7$ which are bound to the SVPs. The system size of the insets, in units of the lattice constant $a$, are: (a) $27 \times 27$; (b) $53 \times 33$; and (c) $63 \times 31$.
	}  
	\label{fig:EnergySpectrum}
\end{figure*}

In the following, we specify our model for the analysis of skyrmion-vortex pairs in such superconductor-ferromagnet heterostructures~\cite{Hals2016, Rex2019b}.
The magnetic structures enter the Hamiltionian effectively via the external magnetic field $\boldsymbol{h}_i$.
For example, for a skyrmion centered at the origin we evaluate the continuous representation
\begin{equation}
\boldsymbol{h}(r, \phi)=h_0\cdot(\cos \Phi \: \sin f(r) , \sin \Phi \: \sin f(r),\cos f(r))^T,
\end{equation} 
with $\Phi= \Phi(\phi)= \mathcal{W}_{\mathrm{sk}} \, \phi + \varphi$ at the corresponding lattice site $i$ with spatial coordinates characterized by radius $r_i$ and azimuthal angle $\phi_i$. Here, $h_0$ is the strength of the proximity-coupled magnetic field, $\mathcal{W}_{\mathrm{sk}}$ is the skyrmion winding number and $\varphi$ is the helicity of the skyrmion.  We approximate the profile function $f(r)$ of the skyrmion by 
$f(r)= 4\, \text{arctan} \bigl(\exp(-r/\lambda)\bigr)$, where $\lambda$ is a parameter describing uniform compression or expansion of the profile, which we choose to match the skyrmion radius to the vortex radius in order to increase the stability of the SVP. 
Skyrmions in thin films~\cite{Leonov2016a} are well approximated by such a profile.

The superconducting order parameter of a single vortex centered at the origin can typically be written as
$\Delta^{\text{Vortex}}_i =\Delta_0(r_i) e^{\mathrm{i}(m \theta_i+\theta_0)}$, where $r_i$ and  $\theta_i$ are respectively the radial distance to and azimuthal angle at lattice site $i$. 
Here, the angle $\theta_0$ (chirality) as well as $m=\pm1$ (vorticity) define the type of the vortex, e.g.\ $\theta_0=0$ and $m=1$ yields a standard outward pointing vortex. Below we discuss how we self-consistently compute the concrete form of the superconducting order parameter induced by the presence of magnetic skyrmions.

In this work, we consider N\'eel skyrmions with $\varphi=0$, which have an attractive interaction with the superconducting vortices such that SVPs form~\cite{Hals2016}. In the next section we show that such SVPs can host Majorana zero modes.


\section{Majorana zero modes at Skyrmion-Vortex Pairs}
\label{sec:energyspectrum1}

Majorana zero modes correspond to excitations, that cost zero energy, between different superconducting ground states in a degenerate ground state manifold. For a given superconducting order parameter $\Delta_i$, the excitations of the system can be obtained using the Bogoliubov-de Gennes (BdG) formalism. This requires rewriting the Hamiltonian, Eq.~\eqref{eq:TightBindingHamiltonian}, as 
$H = \EGS + \tfrac{1}{2}\Psi^{\dagger} \HBdG \Psi$ 
where $\EGS$ is the ground state energy, 
$\Psi = (c_{1\uparrow},c_{1\downarrow},\ldots,c_{N\uparrow},c_{N\downarrow},c_{1\uparrow}^{\dagger},c_{1\downarrow}^{\dagger},\ldots,c_{N\uparrow}^{\dagger},c_{N\downarrow}^{\dagger})^T$, $N$ is the number of lattice sites, and the matrix $\HBdG$ is given in Appendix~\ref{app:HBdG}. Then, a BdG transformation diagonalizes the Hamiltonian into 
$H = \EGS + \sum_n E_n \Gamma_n^{\dagger} \Gamma_n$, where $\Gamma_n = \sum_{i\alpha} ( u_{i\alpha}^{(n)} c_{i\alpha} + v_{i\alpha}^{(n)} c_{i\alpha}^{\dagger} )$ is the annihilation operator of an excitation with energy $E_n$. We adopt the convention that excitations with negative energies are labeled by negative $n$ values~\footnote{Note that the BdG formalism introduces artificially negative energy quasiparticles, but the energies of the physical excitations are positive, and thus are above the ground state.}.
Note that $\Gamma_n$ is a linear superposition of electronic creation and annihilation operators with coefficients determined from the eigenvalue problem
\begin{align}
\label{eq:BdG}\HBdG \, \chi_n = E_n \chi_n, \ \text{where}
\end{align}
$\chi_n = (u_{1\uparrow}^{(n)},u_{1\downarrow}^{(n)},\ldots,u_{N\uparrow}^{(n)},u_{N\downarrow}^{(n)},v_{1\uparrow}^{(n)},v_{1\downarrow}^{(n)},\ldots,v_{N\uparrow}^{(n)},v_{N\downarrow}^{(n)})^T$. 
However, since the spatial dependence of $\Delta_i$ is also unknown, here we solve for the excitation spectrum while self-consistently calculating the superconducting order parameter~\cite{Bjornson2013,Hals2016} 
\begin{equation}
\label{eq:gap}
\Delta_i = \frac{V}{2}
\hspace*{-0.4cm}\sum_{\substack{n>0\\ 
0 \lesssim E_n<\hbar \omega_D}} 
\hspace*{-0.4cm} (\mathrm{i} \sigma_y)_{\alpha\beta}
 v_{i\alpha}^{(n)*}u_{i\beta}^{(n)}\left[2n_F(E_n)-1\right] \,,
\end{equation}   
under the influence of spin-orbit coupling and the magnetic structure, thus revealing the regions in parameter space where Majorana zero modes occur. Here, $n_F(E_n)=(\exp(\frac{E_n}{k_B T})+1)^{-1}$ 
is the Fermi-Dirac distribution, $k_B$ is the Boltzmann constant, $T$ is the temperature, and $V$ characterizes the effective attractive interaction strength between the Cooper pairs. The energy summation is only carried out in a small shell around the Fermi energy, i.e.\ $0 \lesssim E_n<\hbar \omega_D$, where $\omega_D$ is the Debye frequency~\footnote{Note that in our numerical simulations we consider a system of finite size, i.e.\ the Majorana zero modes have a small non-zero energy eigenvalue. Because of the BdG formalism doubling the degrees of freedom in Eq.~\eqref{eq:gap} we only sum over the positive ones.}.
The details of the iteration process can be found in Appendix~\ref{app:energyspectrumrelax}. 
The relaxed superconducting vortex structures are encoded in $\Delta_i$, as sketched in Fig.~\ref{fig:SVP} where the length and direction of the white arrows respectively represent the magnitude and phase of $\Delta_i$.

The particle-hole symmetry inherent to the BdG approach guarantees that $E_{-n} = -E_n$, $u_{i\alpha}^{(n)} = [v_{i\alpha}^{(-n)}]^*$, and that in the special case of vanishing excitation energy $u_{i\alpha} = v_{i\alpha}^{*} = w_{i\alpha}$. Thus, any null eigenvector of the BdG matrix, 
$\HBdG \, \chi = 0$, has the form $\chi = (w_{1\uparrow},w_{1\downarrow},\ldots,w_{N\uparrow},w_{N\downarrow},w_{1\uparrow}^{*},w_{1\downarrow}^{*},\ldots,w_{N\uparrow}^{*},w_{N\downarrow}^{*})^T$. 
The associated excitation annihilation operator 
\begin{align}
\label{eq:gamma}
\gamma = \sum_{i\alpha} \left( w_{i\alpha} c_{i\alpha} +  w_{i\alpha}^* c_{i\alpha}^{\dagger}\right) \,,
\end{align}
corresponds to a Majorana zero mode, namely, $\gamma^{\dagger} = \gamma$. Via the coefficients $w_{i\alpha}$ we will numerically probe the exchange statistics of the Majoranas.

\begin{figure*}[tb]
	\includegraphics[width=\textwidth]{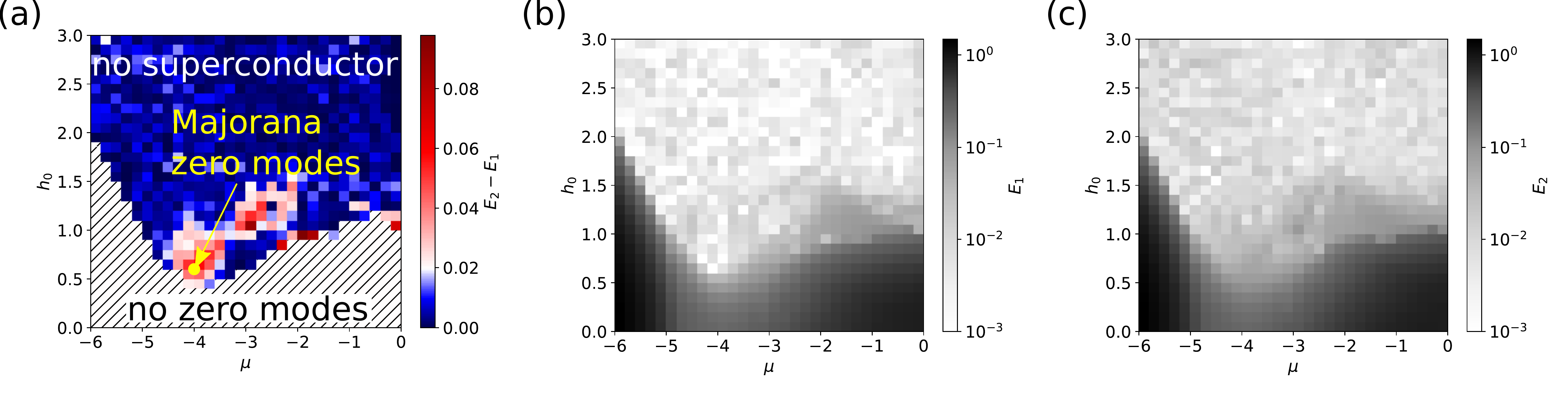}
	\caption{Identifying the topological region in a system with two SVPs. (a) Energy difference $E_2-E_1$ of (b) the lowest positive energy mode $E_1$ and (c) the second lowest positive mode $E_2$, as a function of magnetic field strength $h_0$ and chemical potential $\mu$. For the system to be in a well defined topological region, a finite superconducting gap $E_2-E_1>0$ is required. Majorana zero modes emerge when $E_1 \rightarrow 0$. The yellow dot marks the value of the chemical potential, $\mu=-4$, and magnetic field strength, $h_0=0.7$, which we use to braid Majorana zero modes.
	}
	\label{fig:PhaseDiagram}
\end{figure*}

Figure~\ref{fig:EnergySpectrum} shows the ten lowest positive energy modes and their corresponding negative energy particle-hole partners as a function of the magnetic field strength $h_0$ of a system with (a) two, (b) four, and (c) six SVPs. The physics of an odd number of SVPs is discussed in Appendix~\ref{app:4SVPs}.
The other parameters used to compute the excitation energy spectra were motivated by Refs.~\onlinecite{Hals2016} and \onlinecite{Bjornson2013}, and are summarized in Table~\ref{tab:parameters}~\footnote{Our choice of parameters is further motivated by our results of the analysis of the lowest energy mode for two skyrmion-vortex pairs as function of the chemical potential $\mu$ and the magnetic field strength $h_0$, see Appendix~\ref{app:energyspectrumrelax}}. We find a one-to-one correspondence between the number of SVPs and the number of modes (marked in red for better visibility) whose energies simultaneously approach and become zero at $h_0^{\mathrm{gap}}\approx0.55$---the same for the three systems. Furthermore, for a range of values above $h_0^{\mathrm{gap}}$ the modes whose energy vanished stay tightly close to zero and separate in energy from the other modes. Upon increasing the system size we find that the zero modes become more separated from higher energy modes.
We identify these gapless excitations as Majorana zero modes. Finally, their probability density, obtained using the corresponding coefficients $w_{i\alpha}$, is localized at the SVPs, as shown in the insets of Fig.~\ref{fig:EnergySpectrum}. The remarkable fact that the gap closing for a given model-parameter set is independent of the number of SVPs is consistent with the simultaneous emergence of one Majorana zero mode bound to each SVP.

Characterizing a system with two SVPs is relevant for braiding Majoranas. Thus, in Fig.~\ref{fig:PhaseDiagram}(a) we map out its topological phase diagram in the parameter space $(\mu,h_0)$ to reveal where Majorana zero modes emerge. In the topological region, the lowest energy mode $E_1$, shown in Fig.~\ref{fig:PhaseDiagram}(b), becomes a Majorana zero mode when $E_1 \to 0$ and there is a gap to the next higher energy mode $E_2$, shown in Fig.~\ref{fig:PhaseDiagram}(c). We find the topological region to be around a chemical potential of $\mu=-4$ and a magnetic field strength of about $h_0=0.7$, which we consider in the following.

\begin{table}
	\centering
	\begin{tabular}{c|c|c}
		Parameters & Symbol & Value [$t$] \\
		\hline
		Chemical potential & $\mu$ & $-4$\\
		Spin-Orbit-Coupling & $\alpha_{\text{R}}$ & $0.75$\\
		Thermal energy & $k_B T$ & $0.001$\\
		Debye frequency & $\omega_{D}$ & $100$\\
		Effective attractive interaction & $V$ & $5$\\
	\end{tabular}
\caption{Chosen system parameters in units of the hopping $t$. 
}
\label{tab:parameters}
\end{table}


\section{Braiding of Skyrmion-Vortex Pairs}
\label{sec:braiding}

\begin{figure*}[t]
	\includegraphics[width=\textwidth]{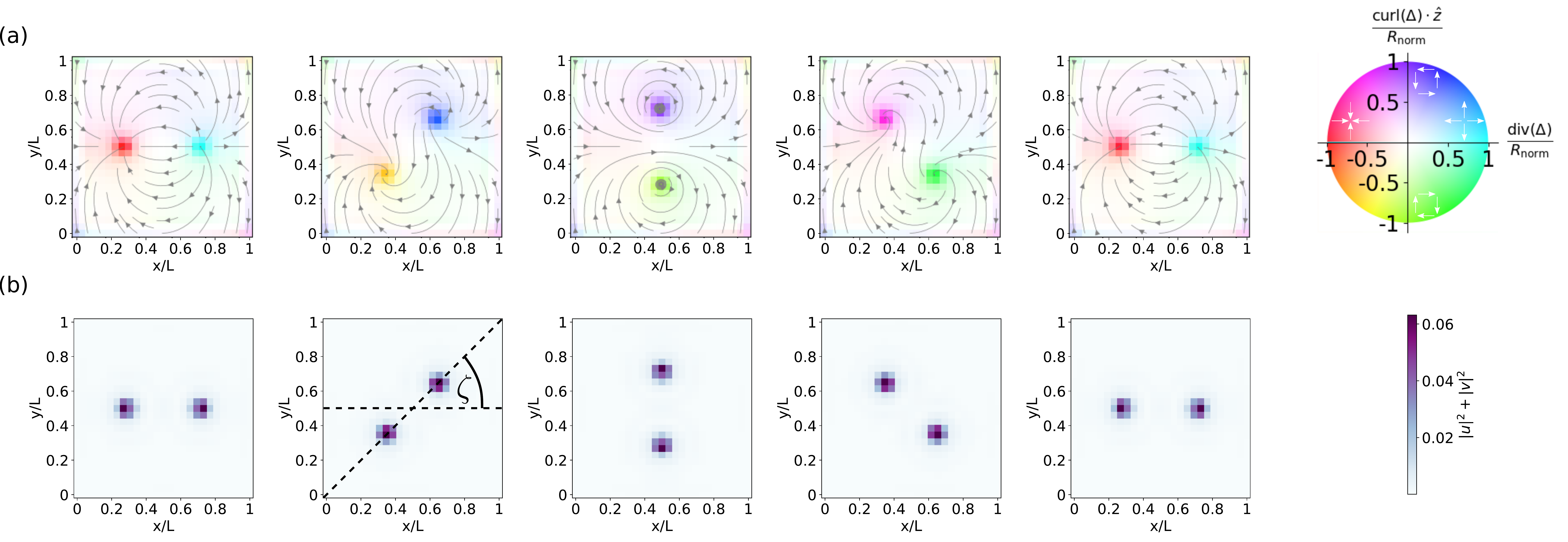}
	\caption{Braiding of two skyrmion-vortex pairs. 
		(a) Evolution of the superconducting order parameter revealing the braiding of the vortices. The color code visualizes the type of the vortex characterized by its divergence and curl, see color wheel on the right. The divergence and curl are normalized by their maximal occurring value $R_{\mathrm{norm}}=0.8$.
		(b) Spatial probability density of the wave functions corresponding to the Majorana zero modes during braiding. The angle $\zeta$, shown in the second panel, parameterizes the brading. In all panels the system size is $L \times L$, with $L=27$ in units of the lattice constant $a$. 
	}  
	\label{fig:2SVP}
\end{figure*}

In this section, we show how we can braid Majorana zero modes by manipulating the SVPs. 
For this we adiabatically change the position of the N\'eel skyrmions and numerically relax the system in each step according to Eqs.~\eqref{eq:BdG} and \eqref{eq:gap}.
Due to the attractive interaction of the N\'eel skyrmion with the superconducting vortex, the latter follows the path of the skyrmion. 
Note that two neighboring vortices need to have opposite chiralities.
For example, an outward pointing vortex must be next to an inward pointing vortex, however two outward pointing vortices cannot be next to each other.
Upon braiding, the vortices change their chirality as exemplified in Fig.~\ref{fig:2SVP}(a) for two SVPs. For example, the inward (outward)-pointing vortex (red and cyan) transforms to a right (left)-handed vortex (green and purple) and then to an outward (inward)-pointing vortex.  
To mimic numerically an infinite sample, we fixed the phase of the superconducting order parameter at the sample boundary. For a large enough sample, we find that this procedure does not affect the zero energy modes.

The braiding of the SVPs also results in the adiabatic motion of the Majorana zero modes bound to the SVPs whose evolution, parameterized by the angle $\zeta \in [0,\pi]$, is shown in 
Fig.~\ref{fig:2SVP}(b). We find that the Majorana zero modes remain well localized upon braiding at the SVPs.
This is also the case when we consider four SVPs, see Appendix~\ref{app:4SVPs}, which we take as a  proof-of-principle demonstration that SVPs hosting Majoranas can provide a scalable platform for topological quantum computing.

In the next section we reveal the non-abelian statistics of the Majorana zero modes during the SVP braiding, thereby providing the evidence that the localized Majorana zero modes obey the statistics of Majorana fermions.


\section{Statistics of Majorana zero modes via SVP braiding}
\label{sec:MZMStatistics}

Calculations in topological quantum computation are foreseen to be performed via braiding operations. Any such operation can be split into a sequence of braids involving only two Majoranas. Let  
$\gamma_j = \sum_{i\alpha} (w_{i\alpha}^{(j)} c_{i\alpha} +  [w_{i\alpha}^{(j)}]^* c_{i\alpha}^{\dagger})$ with $j = 1,2$ be two Majorana operators as introduced in Eq.~\eqref{eq:gamma}. They satisfy the anticommutation relations $\{ \gamma_j, \gamma_{k} \} = 2\delta_{jk}$, provided $\sum_{i\alpha} [w_{i\alpha}^{(j)}]^* w_{i\alpha}^{(k)} = \delta_{jk}$. This can be achieved by maintaining a sufficiently large spatial separation between SVPs during their braiding. Upon exchanging the Majorana zero modes their corresponding operators are expected to transform as $\gamma_1 \to \gamma_2$ and $\gamma_2 \to - \gamma_1$~\cite{Ivanov2001}. This is equivalent to the transformation of the coefficients of the Majorana operators, $w_{i\alpha}^{(1)} \to w_{i\alpha}^{(2)}$ and $w_{i\alpha}^{(2)} \to -w_{i\alpha}^{(1)}$, which we investigate numerically by the adiabatic process of changing $\zeta$ from 0 to $\pi$, i.e., $w_{i\alpha}^{(1)}(\zeta = \pi) = w_{i\alpha}^{(2)}(\zeta = 0)$ and $w_{i\alpha}^{(2)}(\zeta = \pi) = - w_{i\alpha}^{(1)}(\zeta = 0)$. We can succinctly reveal the exchange statistics of the Majorana zero modes by constructing the matrix of overlaps $(M_{12})_{jk} = \sum_{i\alpha} [w_{i\alpha}^{(j)}(\zeta = \pi)]^* w_{i\alpha}^{(k)}(\zeta = 0)$. Our self-consistent calculation for the braiding of any two Majorana zero modes, within numerical error, yields 
\begin{align}\label{eq:Overlaps}
M_{12} =
\begin{pmatrix}
0 & 1\\
-1 & 0
\end{pmatrix}\,.
\end{align}
Crucially, from the orthonormality of the $w_{i\alpha}^{(j)}$ at any given value of $\zeta$ and the off-diagonal elements of $M_{12}$, we conclude that our numerical simulations imply $w_{i\alpha}^{(1)}(\zeta = \pi) = w_{i\alpha}^{(2)}(\zeta = 0)$ and $w_{i\alpha}^{(2)}(\zeta = \pi) = - w_{i\alpha}^{(1)}(\zeta = 0)$. These results explicitly demonstrate the non-abelian statistics of the Majorana zero modes (for details see Appendix~\ref{app:ExchOp}). The diagonal elements of $M_{12}$, serving as a further consistency check, vanish because after exchanging positions, the initial configuration of the $j$-th Majorana zero mode, $w_{i\alpha}^{(j)}(\zeta = 0)$, has no overlap with the final one, $w_{i\alpha}^{(j)}(\zeta = \pi)$.


\section{Scalable Platform for Topological Quantum Computing}
\label{sec:Platform}

\begin{figure*}[t]
   \centering
   \includegraphics[width=0.9\textwidth]{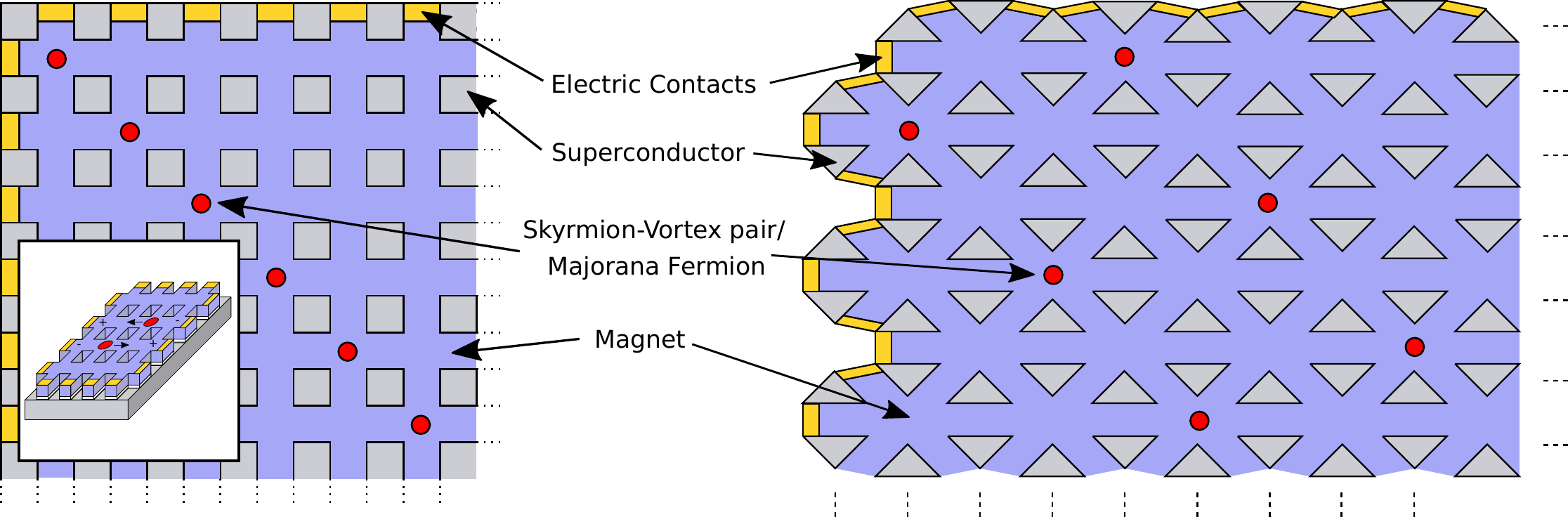}
      \caption{Scalable topological quantum computing platforms based on skyrmion-vortex pairs. Left: topview of a square grid architecture. Inset displays the 3d heterostructure comprising a thin film superconductor (gray) and a square-grid-patterned magnet (violet), separated by an insulating layer (white), where electric contacts (yellow) are used to manipulate and braid SVPs (red). Right: topview of a triangular grid architecture whose diagonal paths allow for denser SVP configurations. Information readout could be done via resistance measurements using the electric contacts after fusing Majorana zero modes by merging SVPs. Adapted from Ref.~\onlinecite{Nothhelferpatent2019}.
      }
   \label{fig:device_with_applied_currents}
\end{figure*}

We have numerically confirmed that Majorana zero modes can be braided via the adiabatic manipulation of SVPs. 
Furthermore, we have also shown that the number of SVPs can be increased with each SVP hosting a Majorana zero mode, as long as their mutual spatial separation is sufficiently large. Crucially, this separation protects the Majorana zero modes from overlapping and hybridizing with each other.
These are important features that suggest we can use SVPs as building blocks of the scalable topological quantum computing platform shown in Fig.~\ref{fig:device_with_applied_currents}.

The architecture of our proposed platform consists of a thin superconducting film on top of which lies a magnetic layer patterned into a grid. 
Electric contacts are placed at the periphery of the magnetic grid to individually address the SVPs. 
An insulating layer located between the magnet and the superconductor protects the latter from electric currents flowing in the former. 
Braiding operations are performed by sequentially applying voltage differences to the appropriate electric contacts. 
Qubits are encoded in the Majorana zero modes and there are several approaches to construct them~\cite{DasSarma2015}. One possible encoding, for example, uses four Majorana zero modes (four SVPs) per qubit, see Appendix~\ref{app:4SVPs}. 
Information can be read out by driving the SVPs toward each other in order to fuse the attached Majorana zero modes and using the electric contacts, measuring the resistance across the merged object~\cite{Nothhelferpatent2019}.

We can estimate the qubit density of our proposed platform by assuming that the tracks of the grid have a width of 100~nm, which is the experimentally reported skyrmion size of recently stabilized SVPs~\cite{Petrovic2021}.
For the specific encoding of four SVPs per qubit, each qubit spans an area of $\sim(8 \times 100$~nm)$^2$, giving a qubit density of $\sim 10^8$~qubit/cm$^2$.
This is much higher than in superconducting qubit systems, whose qubit densities are typically $10^2$~qubit/cm$^2$ as estimated from Ref.~\onlinecite{Neill2018b}.


\section{Discussion and Conclusion}
\label{sec:Conclusion}

Superconductor-ferromagnet heterostructures capable of stabilizing SVPs have been recently identified~\cite{Petrovic2021}. 
Thorough experimental characterization showed that SVPs are formed due to spin-orbit coupling~\cite{Hals2016} as well as stray fields~\cite{Dahir2019}, thus enlarging the engineering possibilities of suitable platforms.
These concrete, already available systems are ideal candidates where our findings on the feasibility of braiding Majorana zero modes attached to SVPs can be tested. 

We have demonstrated that the real-space manipulation of SVPs in two dimensions affords a promising, alternative route to carry out the ``smoking gun'' experiment to reveal the non-abelian statistics of Majorana zero modes. Furthermore, we have proposed a specific device architecture, with SVPs as its building blocks, that can be scaled up to attain the large number of qubits required to reap the anticipated, revolutionary benefits of topological quantum computing.


\section{Acknowledgments}

We thank Achim Rosch, Ilya Eremin, Nico Kerber, Kai Litzius and Niklas Tausendpfund for fruitful discussions.

We acknowledge funding from the German Research Foundation (DFG) Project No. 320163632 (Emmy Noether), Project No. 403233384 (SPP2137 Skyrmionics) and support from the Max Planck Graduate School. 
TM acknowledges funding from the Deutsche Forschungsgemeinschaft via the Emmy Noether Programme ME4844/1-1 (project id 327807255), the Collaborative Research Center SFB 1143 (project id 247310070), and the Cluster of Excellence on Complexity and Topology in Quantum Matter ct.qmat (EXC 2147, project id 390858490). 
This research was supported in part by the National Science Foundation under Grant No. NSF PHY-1748958 and the Research Council of Norway via Grant No. 286889.

\appendix


\section{Bogoliubov-de Gennes Hamiltonian matrix}
\label{app:HBdG}

In the basis chosen in Sec.~\ref{sec:energyspectrum1}, the particle-hole symmetry of the Hamiltonian, Eq.~\eqref{eq:TightBindingHamiltonian} in the main text, manifests in the following block structure of the BdG Hamiltonian matrix 
\begin{equation}
\HBdG = 
    \begin{pmatrix}
        \CalA & \CalB\\
        \CalB^\dagger & -\CalA^*
    \end{pmatrix}
\,,
\end{equation}
where $\CalA = - [\, t\sum_{\braket{i,j}} e_{ij} + \mu\,\mathds{1} ]\otimes\sigma_0 - \sum_i e_{ii}\otimes(\boldsymbol{h}_i\cdot\boldsymbol{\sigma}) + \mathrm{i} \, \alpha_{\text{R}} \sum_{\braket{i,j}} e_{ij}\otimes[\hat{\boldsymbol{z}}\cdot(\boldsymbol{\hat{d}}_{ij}\times\boldsymbol{\sigma})]$ and $\CalB = \sum_i \Delta_i e_{ii}\otimes(\mathrm{i}\sigma_y)$. Here $\sigma_0$ is the $2 \times 2$ identity matrix in spin space. $e_{ij}$ and $\mathds{1}$ are $N \times N$ matrices in position space with $N$ being the number of lattice sites. While $\mathds{1}$ is the identity matrix, the $(i,j)$ entry of $e_{ij}$ is equal to one and all the others vanish.


\section{Relaxation procedure and energy spectrum}
\label{app:energyspectrumrelax}

\begin{figure}[tb]
	\includegraphics[width=0.8\columnwidth]{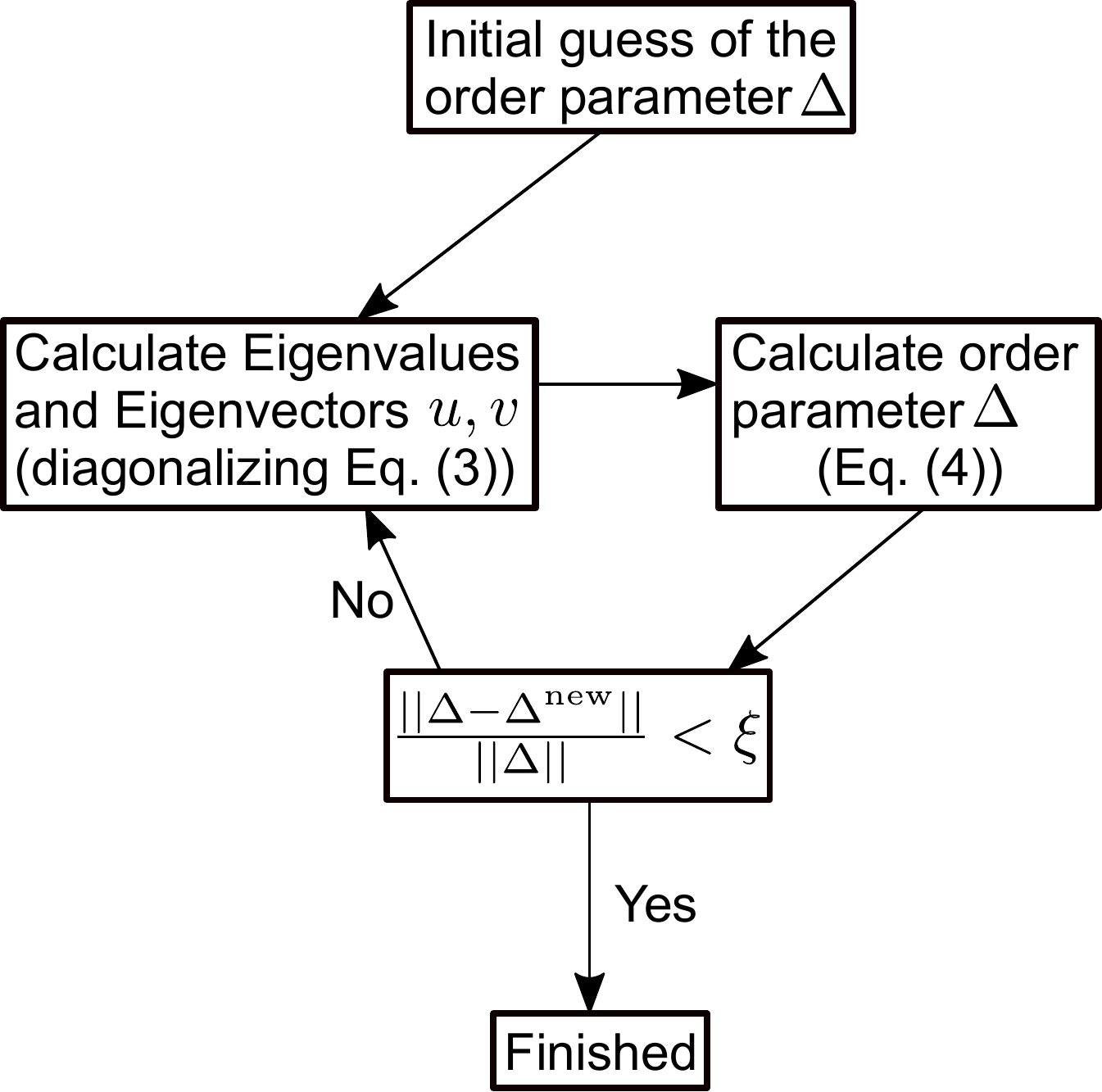}
	\caption{Iteration scheme to calculate the eigensystem of the BdG Hamiltonian matrix, Eq.~\eqref{eq:BdG}, while self-consistently determining the superconducting order parameter, Eq.~\eqref{eq:gap}.
	}  
	\label{fig:SelfConsistentScheme}
\end{figure}

Here we describe in detail the self-consistent calculation of the superconducting order parameter discussed in Sec.~\ref{sec:energyspectrum1}. Figure~\ref{fig:SelfConsistentScheme} depicts the steps of the procedure we have implemented to determine the eigensystem of the BdG Hamiltonian matrix, Eq.~\eqref{eq:BdG} in the main text, while self-consistently computing the superconducting order parameter $\Delta_i$ according to Eq.~\eqref{eq:gap} in the main text. 

In the first step, we make an educated guess for the order parameter. When a single SVP is present in the system, we start with a vortex solution $\Delta^{\text{Vortex}}_i =\Delta_0 e^{\mathrm{i}\theta_i}$ with a constant amplitude $\Delta_0=1$ throughout the sample. For simplicity we define $\theta_i = 0$ at the vortex core, which introduces a discontinuity that will be later smoothed out by the relaxation procedure. For two or more SVPs, we partition the system into sectors each containing a single SVP. The SVP of each sector is initialized as a vortex solution convolved with a Gaussian distribution, thus ensuring that neighboring sectors only have a small mismatch at their interface.

\begin{figure*}[t]
	\includegraphics[width=\textwidth]{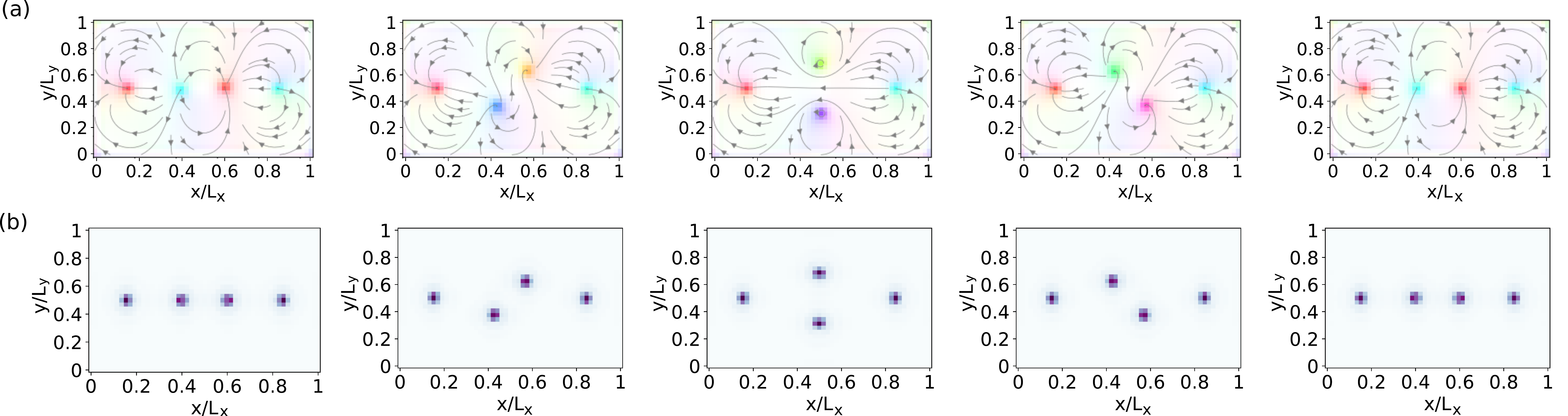}
	\caption{Demonstration of scalability: braiding of two SVPs in a four-SVP set-up. The two SVPs in the middle are braided, leaving the outer ones unchanged. Color codes are as in Fig.~\ref{fig:2SVP}. (a) Evolution of the superconducting order parameter. (b) Spatial probability density of the wave functions corresponding to the Majorana zero modes. In all panels the system size is $L_x \times L_y$, with $L_x=53$ and $L_y=33$ in units of the lattice constant $a$. The original aspect ratio was kept for clarity.}  
	\label{fig:4SVP}
\end{figure*}

In the next step, the eigensystem of Eq.~\eqref{eq:BdG} in the main text is calculated by means of direct diagonalization. With this result we then adjust the solution for the order parameter according to Eq.~\eqref{eq:gap} in the main text. Subsequently, we compare it to the step before and stop if the relative error, $||\Delta - \Delta^{\rm{new}}||/||\Delta||$, is smaller than a cutoff value $\xi$. Note that here we used the Euclidean norm defined as $||\Delta|| = [\, \sum_{i} |\Delta_{i}|^2 \, ]^{1/2}$. For the results shown in Figs.~\ref{fig:EnergySpectrum}-\ref{fig:4SVP} we used $\xi =10^{-4}$.

\section{Numerical calculation of Majorana zero modes}
\label{app:Majoranawavefunctions}

In the calculation of the Majorana zero modes we face the problem that because of the (nearly degenerate) zero energy eigenvalues, any linear combination of the corresponding eigenstates including multiplications with arbitrary phase factors are also solutions to the eigenvalue problem.
To extract the Majorana zero modes we perform a two step procedure, which we demonstrate in the following explicitly for two Majorana zero modes~\footnote{To disentangle multiple zero energy modes we iterate this procedure.}.

Note that we are numerically computing a system of finite size. Therefore, instead of having two identical zero eigenenergies we have a minigap, and thus we can define wave functions with energy $E \gtrsim 0$ and $E \lesssim 0$, because of particle-hole symmetry. 
As we are working in the BdG formalism every eigenstate $\chi$ can be decomposed into an electronic part, $\boldsymbol{u} = (u_{1\uparrow},u_{1\downarrow},\ldots,u_{N\uparrow},u_{N\downarrow})^T$, and a hole part, $\boldsymbol{v} = (v_{1\uparrow},v_{1\downarrow},\ldots,v_{N\uparrow},v_{N\downarrow})^T$, as 
\begin{equation}
\chi = 
    \begin{pmatrix}
	    \boldsymbol{u}\\
		\boldsymbol{v}
    \end{pmatrix}
\,.    
\end{equation}
Additionally, the particle-hole symmetry comes with a gauge choice in the sense that any eigenstate with positive energy
\begin{equation}
\chi_{E \gtrsim 0}=e^{\mathrm{i} \kappa}
    \begin{pmatrix}
	    \boldsymbol{u}\\
		\boldsymbol{v}
    \end{pmatrix}
\,,
\end{equation}
has a corresponding eigenstate, its particle-hole partner, with negative energy,
\begin{equation}
\chi_{E \lesssim 0}=\mathcal{P}\chi_{E \gtrsim 0}=e^{-\mathrm{i} \kappa}
    \begin{pmatrix}
	    \boldsymbol{v}^*\\
		\boldsymbol{u}^*
    \end{pmatrix}
\,.
\end{equation}
Here $\mathcal{P}= \tau_x \mathcal{K}$ is the particle-hole operator, consisting of the operation of exchanging particles and holes combined with complex conjugation.
We have explicitly taken into account an arbitrary, artificial phase factor $e^{\mathrm{i} \kappa}$, which can arise upon diagonalization.
To determine the wave functions and get rid of the arbitrary phase factor, we first write 
\begin{equation}
\label{eq:Eg}
	\chi_{E \gtrsim 0}=  \eta\, \chi_A+ \sqrt{1-\eta^2}\, \chi_B,
\end{equation}
as a linear combination of the computed normalized eigenvectors $\chi_A$ and $\chi_B$, where we vary the parameter $\eta$ to ensure that the wave function has equal weights at both vortex positions. 
Then we compute its particle-hole partner. 

In a second step we compute the Majorana basis~\cite{Wolms2014}
\begin{equation}
	\begin{split}
		\chi_{\text{l}}&= \begin{pmatrix}
			\boldsymbol{u}_{\text{l}}\\
			\boldsymbol{v}_{\text{l}}
		\end{pmatrix}=  \frac{1}{\sqrt{2}}(\tilde{\chi}_{E \gtrsim 0}+\tilde{\chi}_{E \lesssim 0})\\
		&=\frac{1}{\sqrt{2}} 
		\begin{pmatrix}
			e^{\mathrm{i} (\kappa-\beta)} \boldsymbol{u} + e^{-\mathrm{i} (\kappa-\beta)} \boldsymbol{v}^* \\
			e^{\mathrm{i} (\kappa-\beta)} \boldsymbol{v} + e^{-\mathrm{i} (\kappa-\beta)} \boldsymbol{u}^* \\
		\end{pmatrix},
	\end{split}
	\label{eq:psil}
\end{equation}
\begin{equation}
\begin{split}
	\chi_{\text{r}}&=\begin{pmatrix}
		\boldsymbol{u}_{\text{r}}\\
		\boldsymbol{v}_{\text{r}}
	\end{pmatrix}=  \frac{\mathrm{i}}{\sqrt{2}}(\tilde{\chi}_{E \gtrsim 0}-\tilde{\chi}_{E \lesssim 0})\\
    &=\frac{\mathrm{i}}{\sqrt{2}} 
	\begin{pmatrix}
		e^{\mathrm{i} (\kappa-\beta)} \boldsymbol{u} - e^{-\mathrm{i} (\kappa-\beta)} \boldsymbol{v}^* \\
		e^{\mathrm{i} (\kappa-\beta)} \boldsymbol{v} - e^{-\mathrm{i} (\kappa-\beta)} \boldsymbol{u}^* \\
	\end{pmatrix},
\end{split}
\label{eq:psir}
\end{equation}
where we introduced $\tilde{\chi}_{E \gtrsim 0}=e^{-\mathrm{i} \beta}\chi_{E \gtrsim 0}$ and $\tilde{\chi}_{E \lesssim 0}=e^{\mathrm{i} \beta}\chi_{E \lesssim 0}$. We determine $\beta$ such that $\chi_{\text{l}}$ and $\chi_{\text{r}}$ are equally localized at the two vortices. 
We chose the labels `l' and `r' for the initial left and right positions, respectively.
Note that the left/right Majorana zero mode has also a small weight localized at the position of the right/left one, which is exponentially suppressed with the distance between the SVPs.


\section{Scalability}
\label{app:4SVPs}

As shown in Fig.~\ref{fig:EnergySpectrum}, we have performed simulations with more than two SVPs to ensure the scalability of the suggested platform for topological quantum computing. In Fig.~\ref{fig:4SVP} we show four SVPs where we perform the braiding of the inner two, leaving the outer two untouched. 
We do find a similar evolution as in Fig.~\ref{fig:2SVP}, thus demonstrating that the braiding of two SVPs in a background with other Majorana zero modes is feasible.
For this system with four SVPs, the natural extension of the matrix of overlaps, Eq.~\eqref{eq:Overlaps}, is expected to be given by
\begin{align}\label{eq:Overlaps_4SVPs}
M_{1234} =
\begin{pmatrix}
1 & 0 & 0 & 0\\ 
0 & 0 & 1 & 0\\
0 & -1 & 0 & 0\\
0 & 0 & 0 & 1 
\end{pmatrix}\,.
\end{align}
Here 1 and 4 label the outer SVPs, while 2 and 3 the inner ones.

Any two Majorana zero modes give rise to a nonlocal fermion, which defines two so-called parity states: the fermionic state is occupied, $\ket{1}$, or empty, $\ket{0}$.
As mentioned in the main text, four Majorana zero modes can be used to encode one qubit. 
For instance, we can choose to use the parity state of the Majorana zero modes corresponding to SVPs 1 and 2 together with the parity state corresponding to SVPs 3 and 4 to construct the two states of the qubit: $\ket{0,1}$ and $\ket{1,0}$. 
Furthermore, manipulations of the SVPs such as the braiding shown in Fig.~\ref{fig:4SVP} are equivalent to quantum gates acting on this qubit~\cite{Beenakker2020}.

While our results only contain an even number of SVPs, where each Majorana zero mode is located at an SVP, we have also performed simulations with an odd number of SVPs. In these setups we find that all SVPs also bind a Majorana mode, and additionally a Majorana state localizes at the edge of the sample. Adding in another SVP creates a new Majorana zero mode pair. While one Majorana mode localizes at the new SVP, its companion hybridizes with the one that was on the edge, so that again the system has one Majorana zero mode per SVP.


\section{Exchange operator and non-abelian statistics}
\label{app:ExchOp}

Under exchange, two Majorana operators $\gamma_1$ and $\gamma_2$ transform as
\begin{align}
\label{eq:Trafo1}\CalU_{12} \, \gamma_1 \, \CalU_{12}^\dag &= \gamma_2 \,, \\
\label{eq:Trafo2}\CalU_{12} \, \gamma_2 \, \CalU_{12}^\dag &= - \gamma_1 \,,
\end{align}
where the exchange operator is given by~\cite{Ivanov2001}
\begin{align}
\CalU_{12} = e^{-\frac{\pi}{4} \gamma_1 \gamma_2} = \frac{1}{\sqrt{2}} \Big( 1 - \gamma_1 \gamma_2 \Big ) \,.
\end{align}
Note that in Sec.~\ref{sec:MZMStatistics}, we discuss the alternative yet equivalent approach of directly showing the transformation of the Majorana operators, upon an adiabatic exchange, through the corresponding transformation of the coefficients $w_{i\alpha}^{(j)}$.


\bibliography{MyCollection_clean}

\begin{thebibliography}{40}%
\makeatletter
\providecommand \@ifxundefined [1]{%
 \@ifx{#1\undefined}
}%
\providecommand \@ifnum [1]{%
 \ifnum #1\expandafter \@firstoftwo
 \else \expandafter \@secondoftwo
 \fi
}%
\providecommand \@ifx [1]{%
 \ifx #1\expandafter \@firstoftwo
 \else \expandafter \@secondoftwo
 \fi
}%
\providecommand \natexlab [1]{#1}%
\providecommand \enquote  [1]{``#1''}%
\providecommand \bibnamefont  [1]{#1}%
\providecommand \bibfnamefont [1]{#1}%
\providecommand \citenamefont [1]{#1}%
\providecommand \href@noop [0]{\@secondoftwo}%
\providecommand \href [0]{\begingroup \@sanitize@url \@href}%
\providecommand \@href[1]{\@@startlink{#1}\@@href}%
\providecommand \@@href[1]{\endgroup#1\@@endlink}%
\providecommand \@sanitize@url [0]{\catcode `\\12\catcode `\$12\catcode
  `\&12\catcode `\#12\catcode `\^12\catcode `\_12\catcode `\%12\relax}%
\providecommand \@@startlink[1]{}%
\providecommand \@@endlink[0]{}%
\providecommand \url  [0]{\begingroup\@sanitize@url \@url }%
\providecommand \@url [1]{\endgroup\@href {#1}{\urlprefix }}%
\providecommand \urlprefix  [0]{URL }%
\providecommand \Eprint [0]{\href }%
\providecommand \doibase [0]{http://dx.doi.org/}%
\providecommand \selectlanguage [0]{\@gobble}%
\providecommand \bibinfo  [0]{\@secondoftwo}%
\providecommand \bibfield  [0]{\@secondoftwo}%
\providecommand \translation [1]{[#1]}%
\providecommand \BibitemOpen [0]{}%
\providecommand \bibitemStop [0]{}%
\providecommand \bibitemNoStop [0]{.\EOS\space}%
\providecommand \EOS [0]{\spacefactor3000\relax}%
\providecommand \BibitemShut  [1]{\csname bibitem#1\endcsname}%
\let\auto@bib@innerbib\@empty
\bibitem [{\citenamefont {DiVincenzo}(1995)}]{DiVincenzo1995}%
  \BibitemOpen
  \bibfield  {author} {\bibinfo {author} {\bibfnamefont {D.~P.}\ \bibnamefont
  {DiVincenzo}},\ }\href {\doibase 10.1126/science.270.5234.255} {\bibfield
  {journal} {\bibinfo  {journal} {Science}\ }\textbf {\bibinfo {volume}
  {270}},\ \bibinfo {pages} {255} (\bibinfo {year} {1995})}\BibitemShut
  {NoStop}%
\bibitem [{\citenamefont {Landauer}\ \emph {et~al.}(1995)\citenamefont
  {Landauer}, \citenamefont {Welland},\ and\ \citenamefont
  {Gimzewski}}]{Landauer1995}%
  \BibitemOpen
  \bibfield  {author} {\bibinfo {author} {\bibfnamefont {R.}~\bibnamefont
  {Landauer}}, \bibinfo {author} {\bibfnamefont {M.~E.}\ \bibnamefont
  {Welland}}, \ and\ \bibinfo {author} {\bibfnamefont {J.~K.}\ \bibnamefont
  {Gimzewski}},\ }\href {\doibase 10.1098/rsta.1995.0106} {\bibfield  {journal}
  {\bibinfo  {journal} {Philosophical Transactions of the Royal Society of
  London. Series A: Physical and Engineering Sciences}\ }\textbf {\bibinfo
  {volume} {353}},\ \bibinfo {pages} {367} (\bibinfo {year}
  {1995})}\BibitemShut {NoStop}%
\bibitem [{\citenamefont {Pellizzari}\ \emph {et~al.}(1995)\citenamefont
  {Pellizzari}, \citenamefont {Gardiner}, \citenamefont {Cirac},\ and\
  \citenamefont {Zoller}}]{Pellizzari1995}%
  \BibitemOpen
  \bibfield  {author} {\bibinfo {author} {\bibfnamefont {T.}~\bibnamefont
  {Pellizzari}}, \bibinfo {author} {\bibfnamefont {S.~A.}\ \bibnamefont
  {Gardiner}}, \bibinfo {author} {\bibfnamefont {J.~I.}\ \bibnamefont {Cirac}},
  \ and\ \bibinfo {author} {\bibfnamefont {P.}~\bibnamefont {Zoller}},\ }\href
  {\doibase 10.1103/physrevlett.75.3788} {\bibfield  {journal} {\bibinfo
  {journal} {Physical Review Letters}\ }\textbf {\bibinfo {volume} {75}},\
  \bibinfo {pages} {3788} (\bibinfo {year} {1995})}\BibitemShut {NoStop}%
\bibitem [{\citenamefont {Unruh}(1995)}]{Unruh1995}%
  \BibitemOpen
  \bibfield  {author} {\bibinfo {author} {\bibfnamefont {W.~G.}\ \bibnamefont
  {Unruh}},\ }\href {\doibase 10.1103/physreva.51.992} {\bibfield  {journal}
  {\bibinfo  {journal} {Physical Review A}\ }\textbf {\bibinfo {volume} {51}},\
  \bibinfo {pages} {992} (\bibinfo {year} {1995})}\BibitemShut {NoStop}%
\bibitem [{\citenamefont {Pachos}(2009)}]{pachos_2012}%
  \BibitemOpen
  \bibfield  {author} {\bibinfo {author} {\bibfnamefont {J.~K.}\ \bibnamefont
  {Pachos}},\ }\href {\doibase 10.1017/cbo9780511792908} {\emph {\bibinfo
  {title} {Introduction to Topological Quantum Computation}}}\ (\bibinfo
  {publisher} {Cambridge University Press},\ \bibinfo {year} {2009})\ p.\
  \bibinfo {pages} {9780511792908}\BibitemShut {NoStop}%
\bibitem [{\citenamefont {Nayak}\ \emph {et~al.}(2008)\citenamefont {Nayak},
  \citenamefont {Simon}, \citenamefont {Stern}, \citenamefont {Freedman},\ and\
  \citenamefont {Sarma}}]{Nayak2008}%
  \BibitemOpen
  \bibfield  {author} {\bibinfo {author} {\bibfnamefont {C.}~\bibnamefont
  {Nayak}}, \bibinfo {author} {\bibfnamefont {S.~H.}\ \bibnamefont {Simon}},
  \bibinfo {author} {\bibfnamefont {A.}~\bibnamefont {Stern}}, \bibinfo
  {author} {\bibfnamefont {M.}~\bibnamefont {Freedman}}, \ and\ \bibinfo
  {author} {\bibfnamefont {S.~D.}\ \bibnamefont {Sarma}},\ }\href {\doibase
  10.1103/revmodphys.80.1083} {\bibfield  {journal} {\bibinfo  {journal}
  {Reviews of Modern Physics}\ }\textbf {\bibinfo {volume} {80}},\ \bibinfo
  {pages} {1083} (\bibinfo {year} {2008})}\BibitemShut {NoStop}%
\bibitem [{\citenamefont {Kitaev}(2003)}]{Kitaev2003}%
  \BibitemOpen
  \bibfield  {author} {\bibinfo {author} {\bibfnamefont {A.}~\bibnamefont
  {Kitaev}},\ }\href {\doibase 10.1016/s0003-4916(02)00018-0} {\bibfield
  {journal} {\bibinfo  {journal} {Annals of Physics}\ }\textbf {\bibinfo
  {volume} {303}},\ \bibinfo {pages} {2} (\bibinfo {year} {2003})}\BibitemShut
  {NoStop}%
\bibitem [{\citenamefont {Alicea}(2012)}]{Alicea2012}%
  \BibitemOpen
  \bibfield  {author} {\bibinfo {author} {\bibfnamefont {J.}~\bibnamefont
  {Alicea}},\ }\href {\doibase 10.1088/0034-4885/75/7/076501} {\bibfield
  {journal} {\bibinfo  {journal} {Reports on Progress in Physics}\ }\textbf
  {\bibinfo {volume} {75}},\ \bibinfo {pages} {076501} (\bibinfo {year}
  {2012})}\BibitemShut {NoStop}%
\bibitem [{\citenamefont {Leijnse}\ and\ \citenamefont
  {Flensberg}(2012)}]{Leijnse2012}%
  \BibitemOpen
  \bibfield  {author} {\bibinfo {author} {\bibfnamefont {M.}~\bibnamefont
  {Leijnse}}\ and\ \bibinfo {author} {\bibfnamefont {K.}~\bibnamefont
  {Flensberg}},\ }\href {\doibase 10.1088/0268-1242/27/12/124003} {\bibfield
  {journal} {\bibinfo  {journal} {Semiconductor Science and Technology}\
  }\textbf {\bibinfo {volume} {27}},\ \bibinfo {pages} {124003} (\bibinfo
  {year} {2012})}\BibitemShut {NoStop}%
\bibitem [{\citenamefont {Flensberg}\ \emph {et~al.}(2021)\citenamefont
  {Flensberg}, \citenamefont {von Oppen},\ and\ \citenamefont
  {Stern}}]{Flensberg2021}%
  \BibitemOpen
  \bibfield  {author} {\bibinfo {author} {\bibfnamefont {K.}~\bibnamefont
  {Flensberg}}, \bibinfo {author} {\bibfnamefont {F.}~\bibnamefont {von
  Oppen}}, \ and\ \bibinfo {author} {\bibfnamefont {A.}~\bibnamefont {Stern}},\
  }\href {\doibase 10.1038/s41578-021-00336-6} {\bibfield  {journal} {\bibinfo
  {journal} {Nature Reviews Materials}\ }\textbf {\bibinfo {volume} {6}},\
  \bibinfo {pages} {944} (\bibinfo {year} {2021})}\BibitemShut {NoStop}%
\bibitem [{\citenamefont {Kitaev}(2001)}]{Kitaev2000}%
  \BibitemOpen
  \bibfield  {author} {\bibinfo {author} {\bibfnamefont {A.~Y.}\ \bibnamefont
  {Kitaev}},\ }\href {\doibase 10.1070/1063-7869/44/10s/s29} {\bibfield
  {journal} {\bibinfo  {journal} {Physics-Uspekhi}\ }\textbf {\bibinfo {volume}
  {44}},\ \bibinfo {pages} {131} (\bibinfo {year} {2001})}\BibitemShut
  {NoStop}%
\bibitem [{\citenamefont {Lutchyn}\ \emph {et~al.}(2010)\citenamefont
  {Lutchyn}, \citenamefont {Sau},\ and\ \citenamefont
  {Das~Sarma}}]{Lutchyn2010}%
  \BibitemOpen
  \bibfield  {author} {\bibinfo {author} {\bibfnamefont {R.~M.}\ \bibnamefont
  {Lutchyn}}, \bibinfo {author} {\bibfnamefont {J.~D.}\ \bibnamefont {Sau}}, \
  and\ \bibinfo {author} {\bibfnamefont {S.}~\bibnamefont {Das~Sarma}},\ }\href
  {\doibase 10.1103/PhysRevLett.105.077001} {\bibfield  {journal} {\bibinfo
  {journal} {Physical Review Letters}\ }\textbf {\bibinfo {volume} {105}},\
  \bibinfo {pages} {077001} (\bibinfo {year} {2010})}\BibitemShut {NoStop}%
\bibitem [{\citenamefont {Oreg}\ \emph {et~al.}(2010)\citenamefont {Oreg},
  \citenamefont {Refael},\ and\ \citenamefont {von Oppen}}]{Oreg2010}%
  \BibitemOpen
  \bibfield  {author} {\bibinfo {author} {\bibfnamefont {Y.}~\bibnamefont
  {Oreg}}, \bibinfo {author} {\bibfnamefont {G.}~\bibnamefont {Refael}}, \ and\
  \bibinfo {author} {\bibfnamefont {F.}~\bibnamefont {von Oppen}},\ }\href
  {\doibase 10.1103/PhysRevLett.105.177002} {\bibfield  {journal} {\bibinfo
  {journal} {Physical Review Letters}\ }\textbf {\bibinfo {volume} {105}},\
  \bibinfo {pages} {177002} (\bibinfo {year} {2010})}\BibitemShut {NoStop}%
\bibitem [{\citenamefont {Read}\ and\ \citenamefont {Green}(2000)}]{Read2000}%
  \BibitemOpen
  \bibfield  {author} {\bibinfo {author} {\bibfnamefont {N.}~\bibnamefont
  {Read}}\ and\ \bibinfo {author} {\bibfnamefont {D.}~\bibnamefont {Green}},\
  }\href {\doibase 10.1103/physrevb.61.10267} {\bibfield  {journal} {\bibinfo
  {journal} {Physical Review B}\ }\textbf {\bibinfo {volume} {61}},\ \bibinfo
  {pages} {10267} (\bibinfo {year} {2000})}\BibitemShut {NoStop}%
\bibitem [{\citenamefont {Ivanov}(2001)}]{Ivanov2001}%
  \BibitemOpen
  \bibfield  {author} {\bibinfo {author} {\bibfnamefont {D.~A.}\ \bibnamefont
  {Ivanov}},\ }\href {\doibase 10.1103/physrevlett.86.268} {\bibfield
  {journal} {\bibinfo  {journal} {Physical Review Letters}\ }\textbf {\bibinfo
  {volume} {86}},\ \bibinfo {pages} {268} (\bibinfo {year} {2001})}\BibitemShut
  {NoStop}%
\bibitem [{\citenamefont {Fu}\ and\ \citenamefont {Kane}(2008)}]{Fu2008}%
  \BibitemOpen
  \bibfield  {author} {\bibinfo {author} {\bibfnamefont {L.}~\bibnamefont
  {Fu}}\ and\ \bibinfo {author} {\bibfnamefont {C.~L.}\ \bibnamefont {Kane}},\
  }\href {\doibase 10.1103/physrevlett.100.096407} {\bibfield  {journal}
  {\bibinfo  {journal} {Physical Review Letters}\ }\textbf {\bibinfo {volume}
  {100}},\ \bibinfo {pages} {096407} (\bibinfo {year} {2008})}\BibitemShut
  {NoStop}%
\bibitem [{\citenamefont {Lutchyn}\ \emph {et~al.}(2018)\citenamefont
  {Lutchyn}, \citenamefont {Bakkers}, \citenamefont {Kouwenhoven},
  \citenamefont {Krogstrup}, \citenamefont {Marcus},\ and\ \citenamefont
  {Oreg}}]{Lutchyn2018}%
  \BibitemOpen
  \bibfield  {author} {\bibinfo {author} {\bibfnamefont {R.~M.}\ \bibnamefont
  {Lutchyn}}, \bibinfo {author} {\bibfnamefont {E.~P. A.~M.}\ \bibnamefont
  {Bakkers}}, \bibinfo {author} {\bibfnamefont {L.~P.}\ \bibnamefont
  {Kouwenhoven}}, \bibinfo {author} {\bibfnamefont {P.}~\bibnamefont
  {Krogstrup}}, \bibinfo {author} {\bibfnamefont {C.~M.}\ \bibnamefont
  {Marcus}}, \ and\ \bibinfo {author} {\bibfnamefont {Y.}~\bibnamefont
  {Oreg}},\ }\href {\doibase 10.1038/s41578-018-0003-1} {\bibfield  {journal}
  {\bibinfo  {journal} {Nature Reviews Materials}\ }\textbf {\bibinfo {volume}
  {3}},\ \bibinfo {pages} {52} (\bibinfo {year} {2018})}\BibitemShut {NoStop}%
\bibitem [{\citenamefont {Mourik}\ \emph {et~al.}(2012)\citenamefont {Mourik},
  \citenamefont {Zuo}, \citenamefont {Frolov}, \citenamefont {Plissard},
  \citenamefont {Bakkers},\ and\ \citenamefont {Kouwenhoven}}]{Mourik2012c}%
  \BibitemOpen
  \bibfield  {author} {\bibinfo {author} {\bibfnamefont {V.}~\bibnamefont
  {Mourik}}, \bibinfo {author} {\bibfnamefont {K.}~\bibnamefont {Zuo}},
  \bibinfo {author} {\bibfnamefont {S.~M.}\ \bibnamefont {Frolov}}, \bibinfo
  {author} {\bibfnamefont {S.~R.}\ \bibnamefont {Plissard}}, \bibinfo {author}
  {\bibfnamefont {E.~P. A.~M.}\ \bibnamefont {Bakkers}}, \ and\ \bibinfo
  {author} {\bibfnamefont {L.~P.}\ \bibnamefont {Kouwenhoven}},\ }\href
  {\doibase 10.1126/science.1222360} {\bibfield  {journal} {\bibinfo  {journal}
  {Science}\ }\textbf {\bibinfo {volume} {336}},\ \bibinfo {pages} {1003}
  (\bibinfo {year} {2012})}\BibitemShut {NoStop}%
\bibitem [{\citenamefont {Kim}\ \emph {et~al.}(2018)\citenamefont {Kim},
  \citenamefont {Palacio-Morales}, \citenamefont {Posske}, \citenamefont
  {R{\'{o}}zsa}, \citenamefont {Palot{\'{a}}s}, \citenamefont {Szunyogh},
  \citenamefont {Thorwart},\ and\ \citenamefont {Wiesendanger}}]{Kim2018g}%
  \BibitemOpen
  \bibfield  {author} {\bibinfo {author} {\bibfnamefont {H.}~\bibnamefont
  {Kim}}, \bibinfo {author} {\bibfnamefont {A.}~\bibnamefont
  {Palacio-Morales}}, \bibinfo {author} {\bibfnamefont {T.}~\bibnamefont
  {Posske}}, \bibinfo {author} {\bibfnamefont {L.}~\bibnamefont {R{\'{o}}zsa}},
  \bibinfo {author} {\bibfnamefont {K.}~\bibnamefont {Palot{\'{a}}s}}, \bibinfo
  {author} {\bibfnamefont {L.}~\bibnamefont {Szunyogh}}, \bibinfo {author}
  {\bibfnamefont {M.}~\bibnamefont {Thorwart}}, \ and\ \bibinfo {author}
  {\bibfnamefont {R.}~\bibnamefont {Wiesendanger}},\ }\href {\doibase
  10.1126/sciadv.aar5251} {\bibfield  {journal} {\bibinfo  {journal} {Science
  Advances}\ }\textbf {\bibinfo {volume} {4}},\ \bibinfo {pages} {eaar5251}
  (\bibinfo {year} {2018})}\BibitemShut {NoStop}%
\bibitem [{\citenamefont {Alicea}\ \emph {et~al.}(2011)\citenamefont {Alicea},
  \citenamefont {Oreg}, \citenamefont {Refael}, \citenamefont {von Oppen},\
  and\ \citenamefont {Fisher}}]{Alicea2011a}%
  \BibitemOpen
  \bibfield  {author} {\bibinfo {author} {\bibfnamefont {J.}~\bibnamefont
  {Alicea}}, \bibinfo {author} {\bibfnamefont {Y.}~\bibnamefont {Oreg}},
  \bibinfo {author} {\bibfnamefont {G.}~\bibnamefont {Refael}}, \bibinfo
  {author} {\bibfnamefont {F.}~\bibnamefont {von Oppen}}, \ and\ \bibinfo
  {author} {\bibfnamefont {M.~P.~A.}\ \bibnamefont {Fisher}},\ }\href {\doibase
  10.1038/nphys1915} {\bibfield  {journal} {\bibinfo  {journal} {Nature
  Physics}\ }\textbf {\bibinfo {volume} {7}},\ \bibinfo {pages} {412} (\bibinfo
  {year} {2011})}\BibitemShut {NoStop}%
\bibitem [{\citenamefont {Bonderson}\ \emph {et~al.}(2008)\citenamefont
  {Bonderson}, \citenamefont {Freedman},\ and\ \citenamefont
  {Nayak}}]{Bonderson2008}%
  \BibitemOpen
  \bibfield  {author} {\bibinfo {author} {\bibfnamefont {P.}~\bibnamefont
  {Bonderson}}, \bibinfo {author} {\bibfnamefont {M.}~\bibnamefont {Freedman}},
  \ and\ \bibinfo {author} {\bibfnamefont {C.}~\bibnamefont {Nayak}},\ }\href
  {\doibase 10.1103/physrevlett.101.010501} {\bibfield  {journal} {\bibinfo
  {journal} {Physical Review Letters}\ }\textbf {\bibinfo {volume} {101}},\
  \bibinfo {pages} {010501} (\bibinfo {year} {2008})}\BibitemShut {NoStop}%
\bibitem [{\citenamefont {van Heck}\ \emph {et~al.}(2012)\citenamefont {van
  Heck}, \citenamefont {Akhmerov}, \citenamefont {Hassler}, \citenamefont
  {Burrello},\ and\ \citenamefont {Beenakker}}]{vanHeck2012}%
  \BibitemOpen
  \bibfield  {author} {\bibinfo {author} {\bibfnamefont {B.}~\bibnamefont {van
  Heck}}, \bibinfo {author} {\bibfnamefont {A.~R.}\ \bibnamefont {Akhmerov}},
  \bibinfo {author} {\bibfnamefont {F.}~\bibnamefont {Hassler}}, \bibinfo
  {author} {\bibfnamefont {M.}~\bibnamefont {Burrello}}, \ and\ \bibinfo
  {author} {\bibfnamefont {C.~W.~J.}\ \bibnamefont {Beenakker}},\ }\href
  {\doibase 10.1088/1367-2630/14/3/035019} {\bibfield  {journal} {\bibinfo
  {journal} {New Journal of Physics}\ }\textbf {\bibinfo {volume} {14}},\
  \bibinfo {pages} {035019} (\bibinfo {year} {2012})}\BibitemShut {NoStop}%
\bibitem [{\citenamefont {Vijay}\ and\ \citenamefont {Fu}(2016)}]{Vijay2016}%
  \BibitemOpen
  \bibfield  {author} {\bibinfo {author} {\bibfnamefont {S.}~\bibnamefont
  {Vijay}}\ and\ \bibinfo {author} {\bibfnamefont {L.}~\bibnamefont {Fu}},\
  }\href {\doibase 10.1103/physrevb.94.235446} {\bibfield  {journal} {\bibinfo
  {journal} {Physical Review B}\ }\textbf {\bibinfo {volume} {94}},\ \bibinfo
  {pages} {235446} (\bibinfo {year} {2016})}\BibitemShut {NoStop}%
\bibitem [{\citenamefont {Karzig}\ \emph {et~al.}(2017)\citenamefont {Karzig},
  \citenamefont {Knapp}, \citenamefont {Lutchyn}, \citenamefont {Bonderson},
  \citenamefont {Hastings}, \citenamefont {Nayak}, \citenamefont {Alicea},
  \citenamefont {Flensberg}, \citenamefont {Plugge}, \citenamefont {Oreg},
  \citenamefont {Marcus},\ and\ \citenamefont {Freedman}}]{Karzig2017}%
  \BibitemOpen
  \bibfield  {author} {\bibinfo {author} {\bibfnamefont {T.}~\bibnamefont
  {Karzig}}, \bibinfo {author} {\bibfnamefont {C.}~\bibnamefont {Knapp}},
  \bibinfo {author} {\bibfnamefont {R.~M.}\ \bibnamefont {Lutchyn}}, \bibinfo
  {author} {\bibfnamefont {P.}~\bibnamefont {Bonderson}}, \bibinfo {author}
  {\bibfnamefont {M.~B.}\ \bibnamefont {Hastings}}, \bibinfo {author}
  {\bibfnamefont {C.}~\bibnamefont {Nayak}}, \bibinfo {author} {\bibfnamefont
  {J.}~\bibnamefont {Alicea}}, \bibinfo {author} {\bibfnamefont
  {K.}~\bibnamefont {Flensberg}}, \bibinfo {author} {\bibfnamefont
  {S.}~\bibnamefont {Plugge}}, \bibinfo {author} {\bibfnamefont
  {Y.}~\bibnamefont {Oreg}}, \bibinfo {author} {\bibfnamefont {C.~M.}\
  \bibnamefont {Marcus}}, \ and\ \bibinfo {author} {\bibfnamefont {M.~H.}\
  \bibnamefont {Freedman}},\ }\href {\doibase 10.1103/physrevb.95.235305}
  {\bibfield  {journal} {\bibinfo  {journal} {Physical Review B}\ }\textbf
  {\bibinfo {volume} {95}},\ \bibinfo {pages} {235305} (\bibinfo {year}
  {2017})}\BibitemShut {NoStop}%
\bibitem [{\citenamefont {Hals}\ \emph {et~al.}(2016)\citenamefont {Hals},
  \citenamefont {Schecter},\ and\ \citenamefont {Rudner}}]{Hals2016}%
  \BibitemOpen
  \bibfield  {author} {\bibinfo {author} {\bibfnamefont {K.~M.}\ \bibnamefont
  {Hals}}, \bibinfo {author} {\bibfnamefont {M.}~\bibnamefont {Schecter}}, \
  and\ \bibinfo {author} {\bibfnamefont {M.~S.}\ \bibnamefont {Rudner}},\
  }\href {\doibase 10.1103/physrevlett.117.017001} {\bibfield  {journal}
  {\bibinfo  {journal} {Physical Review Letters}\ }\textbf {\bibinfo {volume}
  {117}},\ \bibinfo {pages} {017001} (\bibinfo {year} {2016})}\BibitemShut
  {NoStop}%
\bibitem [{\citenamefont {Rex}\ \emph {et~al.}(2019)\citenamefont {Rex},
  \citenamefont {Gornyi},\ and\ \citenamefont {Mirlin}}]{Rex2019b}%
  \BibitemOpen
  \bibfield  {author} {\bibinfo {author} {\bibfnamefont {S.}~\bibnamefont
  {Rex}}, \bibinfo {author} {\bibfnamefont {I.~V.}\ \bibnamefont {Gornyi}}, \
  and\ \bibinfo {author} {\bibfnamefont {A.~D.}\ \bibnamefont {Mirlin}},\
  }\href {\doibase 10.1103/physrevb.100.064504} {\bibfield  {journal} {\bibinfo
   {journal} {Physical Review B}\ }\textbf {\bibinfo {volume} {100}},\ \bibinfo
  {pages} {064504} (\bibinfo {year} {2019})}\BibitemShut {NoStop}%
\bibitem [{\citenamefont {Nothhelfer}(2019)}]{Nothhelfer2019}%
  \BibitemOpen
  \bibfield  {author} {\bibinfo {author} {\bibfnamefont {J.}~\bibnamefont
  {Nothhelfer}},\ }\href@noop {} {\bibfield  {journal} {\bibinfo  {journal}
  {Master thesis, Johannes Gutenb. Univ. Mainz}\ } (\bibinfo {year}
  {2019})}\BibitemShut {NoStop}%
\bibitem [{\citenamefont {Nothhelfer}\ \emph {et~al.}(2019)\citenamefont
  {Nothhelfer}, \citenamefont {Hals}, \citenamefont {Everschor-Sitte},\ and\
  \citenamefont {Rizzi}}]{Nothhelferpatent2019}%
  \BibitemOpen
  \bibfield  {author} {\bibinfo {author} {\bibfnamefont {J.}~\bibnamefont
  {Nothhelfer}}, \bibinfo {author} {\bibfnamefont {K.~M.~D.}\ \bibnamefont
  {Hals}}, \bibinfo {author} {\bibfnamefont {K.}~\bibnamefont
  {Everschor-Sitte}}, \ and\ \bibinfo {author} {\bibfnamefont {M.}~\bibnamefont
  {Rizzi}},\ }\href
  {https://data.epo.org/publication-server/pdf-document?pn=3751472&ki=A1&cc=EP&pd=20201216}
  {\bibfield  {journal} {\bibinfo  {journal} {European Patent Application
  EP3751472A1 / International Patent Application WO2020EP66120}\ } (\bibinfo
  {year} {2019})}\BibitemShut {NoStop}%
\bibitem [{\citenamefont {Björnson}\ and\ \citenamefont
  {Black-Schaffer}(2013)}]{Bjornson2013}%
  \BibitemOpen
  \bibfield  {author} {\bibinfo {author} {\bibfnamefont {K.}~\bibnamefont
  {Björnson}}\ and\ \bibinfo {author} {\bibfnamefont {A.~M.}\ \bibnamefont
  {Black-Schaffer}},\ }\href {\doibase 10.1103/physrevb.88.024501} {\bibfield
  {journal} {\bibinfo  {journal} {Physical Review B}\ }\textbf {\bibinfo
  {volume} {88}},\ \bibinfo {pages} {024501} (\bibinfo {year}
  {2013})}\BibitemShut {NoStop}%
\bibitem [{\citenamefont {Leonov}\ \emph {et~al.}(2016)\citenamefont {Leonov},
  \citenamefont {Monchesky}, \citenamefont {Romming}, \citenamefont {Kubetzka},
  \citenamefont {Bogdanov},\ and\ \citenamefont {Wiesendanger}}]{Leonov2016a}%
  \BibitemOpen
  \bibfield  {author} {\bibinfo {author} {\bibfnamefont {A.~O.}\ \bibnamefont
  {Leonov}}, \bibinfo {author} {\bibfnamefont {T.~L.}\ \bibnamefont
  {Monchesky}}, \bibinfo {author} {\bibfnamefont {N.}~\bibnamefont {Romming}},
  \bibinfo {author} {\bibfnamefont {A.}~\bibnamefont {Kubetzka}}, \bibinfo
  {author} {\bibfnamefont {A.~N.}\ \bibnamefont {Bogdanov}}, \ and\ \bibinfo
  {author} {\bibfnamefont {R.}~\bibnamefont {Wiesendanger}},\ }\href {\doibase
  10.1088/1367-2630/18/6/065003} {\bibfield  {journal} {\bibinfo  {journal}
  {New Journal of Physics}\ }\textbf {\bibinfo {volume} {18}},\ \bibinfo
  {pages} {065003} (\bibinfo {year} {2016})}\BibitemShut {NoStop}%
\bibitem [{Note1()}]{Note1}%
  \BibitemOpen
  \bibinfo {note} {Note that the BdG formalism introduces artificially negative
  energy quasiparticles, but the energies of the physical excitations are
  positive, and thus are above the ground state.}\BibitemShut {Stop}%
\bibitem [{Note2()}]{Note2}%
  \BibitemOpen
  \bibinfo {note} {Note that in our numerical simulations we consider a system
  of finite size, i.e.\ the Majorana zero modes have a small non-zero energy
  eigenvalue. Because of the BdG formalism doubling the degrees of freedom in
  Eq.~\protect \textup {\hbox {\mathsurround \z@ \protect \normalfont
  (\ignorespaces \ref {eq:gap}\unskip \@@italiccorr )}} we only sum over the
  positive ones.}\BibitemShut {Stop}%
\bibitem [{Note3()}]{Note3}%
  \BibitemOpen
  \bibinfo {note} {Our choice of parameters is further motivated by our results
  of the analysis of the lowest energy mode for two skyrmion-vortex pairs as
  function of the chemical potential $\mu $ and the magnetic field strength
  $h_0$, see Appendix~\ref {app:energyspectrumrelax}}\BibitemShut {NoStop}%
\bibitem [{\citenamefont {Sarma}\ \emph {et~al.}(2015)\citenamefont {Sarma},
  \citenamefont {Freedman},\ and\ \citenamefont {Nayak}}]{DasSarma2015}%
  \BibitemOpen
  \bibfield  {author} {\bibinfo {author} {\bibfnamefont {S.~D.}\ \bibnamefont
  {Sarma}}, \bibinfo {author} {\bibfnamefont {M.}~\bibnamefont {Freedman}}, \
  and\ \bibinfo {author} {\bibfnamefont {C.}~\bibnamefont {Nayak}},\ }\href
  {\doibase 10.1038/npjqi.2015.1} {\bibfield  {journal} {\bibinfo  {journal}
  {npj Quantum Information}\ }\textbf {\bibinfo {volume} {1}},\ \bibinfo
  {pages} {15001} (\bibinfo {year} {2015})}\BibitemShut {NoStop}%
\bibitem [{\citenamefont {Petrovi{\'{c}}}\ \emph {et~al.}(2021)\citenamefont
  {Petrovi{\'{c}}}, \citenamefont {Raju}, \citenamefont {Tee}, \citenamefont
  {Louat}, \citenamefont {Maggio-Aprile}, \citenamefont {Menezes},
  \citenamefont {Wyszy{\'{n}}ski}, \citenamefont {Duong}, \citenamefont
  {Reznikov}, \citenamefont {Renner}, \citenamefont {Milo{\v{s}}evi{\'{c}}},\
  and\ \citenamefont {Panagopoulos}}]{Petrovic2021}%
  \BibitemOpen
  \bibfield  {author} {\bibinfo {author} {\bibfnamefont {A.}~\bibnamefont
  {Petrovi{\'{c}}}}, \bibinfo {author} {\bibfnamefont {M.}~\bibnamefont
  {Raju}}, \bibinfo {author} {\bibfnamefont {X.}~\bibnamefont {Tee}}, \bibinfo
  {author} {\bibfnamefont {A.}~\bibnamefont {Louat}}, \bibinfo {author}
  {\bibfnamefont {I.}~\bibnamefont {Maggio-Aprile}}, \bibinfo {author}
  {\bibfnamefont {R.}~\bibnamefont {Menezes}}, \bibinfo {author} {\bibfnamefont
  {M.}~\bibnamefont {Wyszy{\'{n}}ski}}, \bibinfo {author} {\bibfnamefont
  {N.}~\bibnamefont {Duong}}, \bibinfo {author} {\bibfnamefont
  {M.}~\bibnamefont {Reznikov}}, \bibinfo {author} {\bibfnamefont
  {C.}~\bibnamefont {Renner}}, \bibinfo {author} {\bibfnamefont
  {M.}~\bibnamefont {Milo{\v{s}}evi{\'{c}}}}, \ and\ \bibinfo {author}
  {\bibfnamefont {C.}~\bibnamefont {Panagopoulos}},\ }\href {\doibase
  10.1103/physrevlett.126.117205} {\bibfield  {journal} {\bibinfo  {journal}
  {Physical Review Letters}\ }\textbf {\bibinfo {volume} {126}},\ \bibinfo
  {pages} {117205} (\bibinfo {year} {2021})}\BibitemShut {NoStop}%
\bibitem [{\citenamefont {Neill}\ \emph {et~al.}(2018)\citenamefont {Neill},
  \citenamefont {Roushan}, \citenamefont {Kechedzhi}, \citenamefont {Boixo},
  \citenamefont {Isakov}, \citenamefont {Smelyanskiy}, \citenamefont {Megrant},
  \citenamefont {Chiaro}, \citenamefont {Dunsworth}, \citenamefont {Arya},
  \citenamefont {Barends}, \citenamefont {Burkett}, \citenamefont {Chen},
  \citenamefont {Chen}, \citenamefont {Fowler}, \citenamefont {Foxen},
  \citenamefont {Giustina}, \citenamefont {Graff}, \citenamefont {Jeffrey},
  \citenamefont {Huang}, \citenamefont {Kelly}, \citenamefont {Klimov},
  \citenamefont {Lucero}, \citenamefont {Mutus}, \citenamefont {Neeley},
  \citenamefont {Quintana}, \citenamefont {Sank}, \citenamefont {Vainsencher},
  \citenamefont {Wenner}, \citenamefont {White}, \citenamefont {Neven},\ and\
  \citenamefont {Martinis}}]{Neill2018b}%
  \BibitemOpen
  \bibfield  {author} {\bibinfo {author} {\bibfnamefont {C.}~\bibnamefont
  {Neill}}, \bibinfo {author} {\bibfnamefont {P.}~\bibnamefont {Roushan}},
  \bibinfo {author} {\bibfnamefont {K.}~\bibnamefont {Kechedzhi}}, \bibinfo
  {author} {\bibfnamefont {S.}~\bibnamefont {Boixo}}, \bibinfo {author}
  {\bibfnamefont {S.~V.}\ \bibnamefont {Isakov}}, \bibinfo {author}
  {\bibfnamefont {V.}~\bibnamefont {Smelyanskiy}}, \bibinfo {author}
  {\bibfnamefont {A.}~\bibnamefont {Megrant}}, \bibinfo {author} {\bibfnamefont
  {B.}~\bibnamefont {Chiaro}}, \bibinfo {author} {\bibfnamefont
  {A.}~\bibnamefont {Dunsworth}}, \bibinfo {author} {\bibfnamefont
  {K.}~\bibnamefont {Arya}}, \bibinfo {author} {\bibfnamefont {R.}~\bibnamefont
  {Barends}}, \bibinfo {author} {\bibfnamefont {B.}~\bibnamefont {Burkett}},
  \bibinfo {author} {\bibfnamefont {Y.}~\bibnamefont {Chen}}, \bibinfo {author}
  {\bibfnamefont {Z.}~\bibnamefont {Chen}}, \bibinfo {author} {\bibfnamefont
  {A.}~\bibnamefont {Fowler}}, \bibinfo {author} {\bibfnamefont
  {B.}~\bibnamefont {Foxen}}, \bibinfo {author} {\bibfnamefont
  {M.}~\bibnamefont {Giustina}}, \bibinfo {author} {\bibfnamefont
  {R.}~\bibnamefont {Graff}}, \bibinfo {author} {\bibfnamefont
  {E.}~\bibnamefont {Jeffrey}}, \bibinfo {author} {\bibfnamefont
  {T.}~\bibnamefont {Huang}}, \bibinfo {author} {\bibfnamefont
  {J.}~\bibnamefont {Kelly}}, \bibinfo {author} {\bibfnamefont
  {P.}~\bibnamefont {Klimov}}, \bibinfo {author} {\bibfnamefont
  {E.}~\bibnamefont {Lucero}}, \bibinfo {author} {\bibfnamefont
  {J.}~\bibnamefont {Mutus}}, \bibinfo {author} {\bibfnamefont
  {M.}~\bibnamefont {Neeley}}, \bibinfo {author} {\bibfnamefont
  {C.}~\bibnamefont {Quintana}}, \bibinfo {author} {\bibfnamefont
  {D.}~\bibnamefont {Sank}}, \bibinfo {author} {\bibfnamefont {A.}~\bibnamefont
  {Vainsencher}}, \bibinfo {author} {\bibfnamefont {J.}~\bibnamefont {Wenner}},
  \bibinfo {author} {\bibfnamefont {T.~C.}\ \bibnamefont {White}}, \bibinfo
  {author} {\bibfnamefont {H.}~\bibnamefont {Neven}}, \ and\ \bibinfo {author}
  {\bibfnamefont {J.~M.}\ \bibnamefont {Martinis}},\ }\href {\doibase
  10.1126/science.aao4309} {\bibfield  {journal} {\bibinfo  {journal}
  {Science}\ }\textbf {\bibinfo {volume} {360}},\ \bibinfo {pages} {195}
  (\bibinfo {year} {2018})}\BibitemShut {NoStop}%
\bibitem [{\citenamefont {Dahir}\ \emph {et~al.}(2019)\citenamefont {Dahir},
  \citenamefont {Volkov},\ and\ \citenamefont {Eremin}}]{Dahir2019}%
  \BibitemOpen
  \bibfield  {author} {\bibinfo {author} {\bibfnamefont {S.~M.}\ \bibnamefont
  {Dahir}}, \bibinfo {author} {\bibfnamefont {A.~F.}\ \bibnamefont {Volkov}}, \
  and\ \bibinfo {author} {\bibfnamefont {I.~M.}\ \bibnamefont {Eremin}},\
  }\href {\doibase 10.1103/physrevlett.122.097001} {\bibfield  {journal}
  {\bibinfo  {journal} {Physical Review Letters}\ }\textbf {\bibinfo {volume}
  {122}},\ \bibinfo {pages} {097001} (\bibinfo {year} {2019})}\BibitemShut
  {NoStop}%
\bibitem [{Note4()}]{Note4}%
  \BibitemOpen
  \bibinfo {note} {To disentangle multiple zero energy modes we iterate this
  procedure.}\BibitemShut {Stop}%
\bibitem [{\citenamefont {Wölms}\ \emph {et~al.}(2014)\citenamefont {Wölms},
  \citenamefont {Stern},\ and\ \citenamefont {Flensberg}}]{Wolms2014}%
  \BibitemOpen
  \bibfield  {author} {\bibinfo {author} {\bibfnamefont {K.}~\bibnamefont
  {Wölms}}, \bibinfo {author} {\bibfnamefont {A.}~\bibnamefont {Stern}}, \
  and\ \bibinfo {author} {\bibfnamefont {K.}~\bibnamefont {Flensberg}},\ }\href
  {\doibase 10.1103/physrevlett.113.246401} {\bibfield  {journal} {\bibinfo
  {journal} {Physical Review Letters}\ }\textbf {\bibinfo {volume} {113}},\
  \bibinfo {pages} {246401} (\bibinfo {year} {2014})}\BibitemShut {NoStop}%
\bibitem [{\citenamefont {Beenakker}(2020)}]{Beenakker2020}%
  \BibitemOpen
  \bibfield  {author} {\bibinfo {author} {\bibfnamefont {C.~W.~J.}\
  \bibnamefont {Beenakker}},\ }\href {\doibase
  10.21468/SciPostPhysLectNotes.15} {\bibfield  {journal} {\bibinfo  {journal}
  {SciPost Phys. Lect. Notes}\ ,\ \bibinfo {pages} {15}} (\bibinfo {year}
  {2020})}\BibitemShut {NoStop}%
\end{thebibliography}%

\end{document}